# Fragile phase stability in (1-$x$)Pb(Mg$_{1/3}$Nb$_{2/3}$O$_3$)-$x$PbTiO$_3$ crystals: A comparisons of [001] and [110] field-cooled phase diagrams

Hu Cao, Jiefang Li and D. Viehland

*Dept. of Materials Science and Engineering, Virginia Tech, Blacksburg, VA 24061*

Guangyong Xu

*Condensed Matter Physics and Materials Science Department, Brookhaven National Laboratory, Upton, NY 11973*

Phase diagrams of [001] and [110] field-cooled (FC) (1-$x$)Pb(Mg$_{1/3}$Nb$_{2/3}$O$_3$)-$x$PbTiO$_3$ or PMN-$x$PT (0.15≤$x$≤0.38) crystals have been constructed, based on high-resolution x-ray diffraction data. Comparisons reveal several interesting findings. First, a region of abnormal thermal expansion ($c≠a$) above the dielectric maximum was found, whose stability range extended to higher temperatures by application of electric field (E). Second, the rhombohedral (R) phase of the ZFC state was replaced by a monoclinic M$_A$ in the [001] FC diagram, but with monoclinic M$_B$ in the [110] FC. Third, the monoclinic M$_C$ phase in ZFC and [001] FC diagram was replaced by an orthorhombic (O) phase in the [110] FC. Finally, in the [001] FC diagram, the phase boundary between tetragonal (T) and M$_A$ was extended to lower PT contents ($x$=0.25); whereas in the [110] FC diagram, this extended region was entirely replaced by the O phase. *These results clearly demonstrate that the phase stability of PMN-xPT crystals is quite fragile – depending not only on modest changes in E, but also on the direction along which that E is applied.*

## I. Introduction

Solid solutions of (1-$x$)Pb(Mg$_{1/3}$Nb$_{2/3}$O$_3$)-$x$PbTiO$_3$ (PMN-$x$PT) and (1-$x$)Pb(Zn$_{1/3}$Nb$_{2/3}$O$_3$)-$x$PbTiO$_3$ (PZN-$x$PT), have attracted much interests as high performance piezoelectric actuator and transducer materials[1]. For example, (001)-oriented PMN-0.33PT crystals, which lies inside of a morphotropic phase boundary (MPB), have the highest piezoelectric ($d_{33}$~2500pC/N) and electromechanical coupling ($k_{33}$~94%) coefficients[2]. Historically, the high electromechanical properties of Pb(Zr$_x$,Ti$_{1-x}$)O$_3$ (PZT) ceramics were attributed to the nearly-vertical MPB[3] between rhombohedral (R) and tetragonal (T) ferroelectric phases, resulting in phase coexistence. Park and Shrout conjectured that the exceptional electromechanical properties of oriented PMN-$x$PT and PZN-$x$PT crystals were rather due to a R→T phase transition induced by an applied electric field (E)[1,4].

More recently, various intermediate monoclinic phases that structurally 'bridge' the R and T ones across the boundary have been reported in PZT ceramics[5-7], and subsequently in PZN-$x$PT[8-12] and PMN-$x$PT[12-17] crystals – i.e., the MPB is not so vertical and sharp. Monoclinicity may be important in that it allows the polarization vector to be unconstrained within a plane[18], rather than constricted to a particular crystallographic axis as for the higher symmetry R, T, or orthorhombic (O) phases, as shown in Figure 1. Two types of monoclinic distortions $M_A$ and $M_C$ have been reported, which correspond to space groups *Cm* and *Pm*, respectively. The $M_A$ unit cell has a unique $b_m$ axis along the [110] direction, and is doubled and rotated 45º about the *c*-axis, with respect to the pseudocubic cell; whereas, the $M_C$ unit cell is primitive having a unique $b_m$ axis that is oriented along the pseudocubic [010]. Recently, a monoclinic $M_B$ phase has been reported[17]: although both the $M_A$ and $M_B$ phases belong to the *Cm* space group, the difference lies in the magnitudes of the components of the polarization[19] corresponding to the pseudocubic cell: for the $M_A$ phase, $P_x=P_y<P_z$, whereas for the $M_B$ phase, $P_x=P_y>P_z$. In addition, an O ferroelectric phase has been reported to be induced by E in (001) PZN-0.08PT[8] crystals, and by a field applied along (110) in PMN-0.30PT[17]. This O phase is the limiting case of a $M_C$ phase, which can be considered as $a_m=c_m$, similar to that of $BaTiO_3$[20].

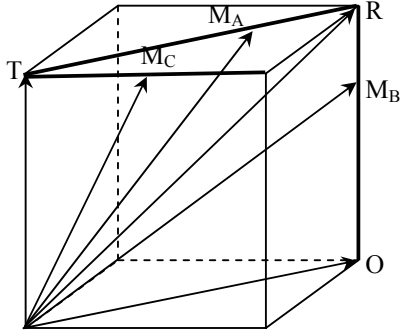

Figure 1. Illustration of rotation of polarization vectors in monoclinic perovskite unit cells. The thick lines represent the paths followed by the end of the polarization vector in the $M_A$ phase where the polarization rotates in a plane between rhombohedral (R) and tetragonal (T), $M_B$ phase where the polarization rotates between R and orthorhombic (O) phases, and $M_C$ phase where the polarization rotates between O and T phases. The $M_A$, $M_B$ and $M_C$ notation is adopted following Vanderbilt and Cohen (Ref. 18).

Recent diffraction experiments under electric field (E) have shown that the phase stability of PMN-*x*PT and PZN-*x*PT is dependent upon the electrical history of the crystal. Neutron studies of PZN-0.08PT crystals with E//[001][11], have shown a C→T→$M_C$ phase sequence in the field-cooled (FC) condition, but a R→$M_A$→$M_C$→T one with increasing E beginning from the zero-field-cooled (ZFC) state. Similarly, neutron and x-ray diffraction investigations[15] of PMN-0.30PT have established a C→T→$M_C$→$M_A$ sequence in the E//[001] FC condition, but a R→$M_A$→$M_C$→T one with increasing E beginning from the ZFC. Dielectric property studies of PMN-0.33PT[21] crystals with E//[110] have reported a mestastable O phase, bridging T and R ones, over a narrow temperature range in the FC condition. Polarized light microscopy (PLM) indicated that this O phase was a single domain one. A recent structural study of PMN-0.30PT with E//[110][17] has unambiguously shown a C→T→O→$M_B$ sequence in the FC condition, but a R→$M_B$→O one with increasing E beginning from the ZFC at 300K. These prior studies clearly evidence that the phase stability of PMN-*x*PT crystals is altered by electrical history and by crystallographic direction along which that E is applied. However, a systematic investigation of the phase stability has not yet been performed over a wide compositional range of *x*. Thus, understanding of how the phase diagram is altered by the direction along which E is applied is limited.

Here, we report a high-resolution x-ray diffraction study of [001] and [110] FC PMN-*x*PT crystals of numerous compositions, both near and away from the MPB. Our findings are summarized in Figure 2. We have determined how the stability regions of the various intermediate phases are altered by E and the direction along which it is applied. *The results demonstrate that the phase diagram of PMN-xPT crystals is quite fragile – depending not only on modest changes in E, but also on* *the direction along which that E is applied.*

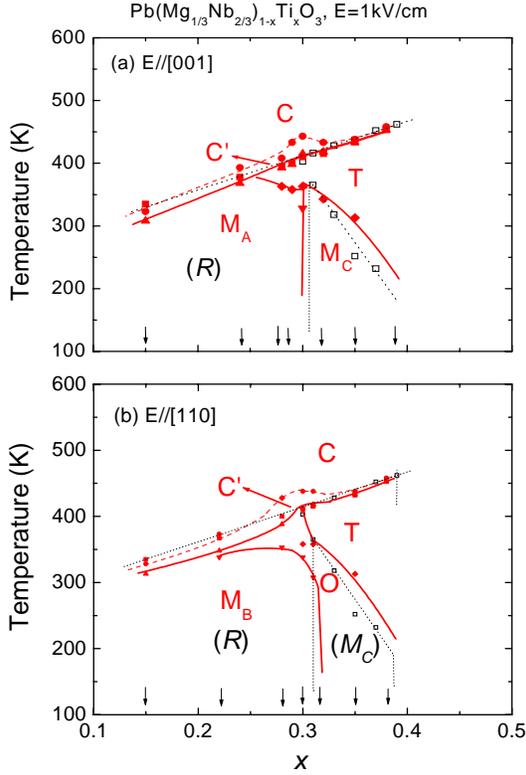

Figure 2. Modified phase diagrams of (a) [001] and (b) [110] electric field cooled PMN-*x*PT crystals. The dotted lines and open square signs were based on prior studies by Noheda *et al.* (Ref. 14). The bracketed italic R represents the rhomboheral phase of the zero-field-cooled condition. The solid square signs represent the temperature of the dielectric maximum ($T_M$). The C' phase below the upper dashed curve was determined by a region of abnormal thermal expansion. Solid curves drawn through these data points are only for guide of eyes.

## II. Experiment Procedure

Single crystals of PMN-*x*PT (*x*=0.38, 0.35, 0.32, 0.30, 0.28, 0.27, 0.24, 0.15) with dimension of 3×3×3 mm³ were obtained from HC Materials (Urbana, IL), and were grown by a top-seeded modified Bridgman method. Two kinds of cubes were cut along the pseudocubic (001)/(100)/(010) and $(110)/(1\bar{1}1)/(\bar{1}12)$ planes, and were polished to 0.25μm. Gold electrodes were deposited by sputtering. Temperature dependent dielectric constant measurements were performed using a multi-frequency LCR meter (HP 4284A) under various E. The XRD studies were performed using a Philips MPD high-resolution system equipped with a two bounce hybrid monochromator, an open 3-circle Eulerian cradle, and a doomed hot-stage. A Ge (220)-cut crystal was used as an analyzer, which had an angular resolution of 0.0068°. The x-ray wavelength was that of $Cu_{K\alpha}$=1.5406Å, and the x-ray generator was operated at 45kV and 40mA. The penetration depth in the samples was on the order of 10 microns. For (001)-field-cooled PMN-xPT crystals, we performed mesh scans around the (002) and (200) Bragg reflections in the (H,0,L) zone, defined by the [001] and [100] vectors; and about the (220) reflection in the scattering (H,H,L) zone, defined by the [110] and [001] vectors. For (110)-fielded PMN-*x*PT, the domain structure is more complicated: we performed mesh scans around the (002) reflection in the (H,H,L), defined by the [001] and [110] vectors; the (220) and $(2\bar{2}0)$ reflections in the scattering zone defined by the [110] and $[1\bar{1}0]$ vectors; and the (200) in the (H0L) zone, defined by the [100] and [001] vectors. Each measurement cycle was begun by heating up to 550K to depole the crystal, and measurements subsequently taken on cooling. In this study we fixed the reciprocal lattice unit (or l rlu) a*=2π/a=1.560Å⁻¹. All mesh scans of PMN-*x*PT shown in this study were plotted in reference to this reciprocal unit.

## III. Results

**[001] electric field cooled PMN-*x*PT**

Figure 3 shows the temperature evolution of the lattice parameters for [001] FC PMN-*x*PT crystals whose composition is to the left of the MPB: x=0.15, 0.24, 0.27, and 0.28. Previously, we have reported those for compositions near and to the right of the MPB, please see the reference given for the respective composition: x=0.30[15], 0.32[16], and 0.35[22].

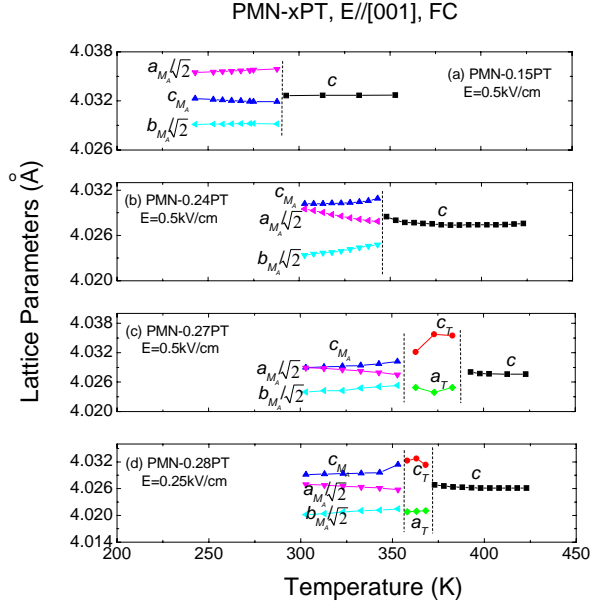

Figure 3. Lattice parameters as a function of temperature for [001] field cooled PMN-xPT crystals: (a) PMN-0.15PT, E=0.5kv/cm; (b) PMN-0.24PT, E=0.5kv/cm; (c) PMN-0.27PT, E=0.5kv/cm; and (d) PMN-0.28PT under E=0.25kv/cm.

For PMN-0.15PT and PMN-0.24PT, only a single phase transition was observed on cooling under a field of E=0.5kV/cm. However, there was an important difference between the temperature evolutions of the lattice parameters for these two compositions. Specifically for x=0.15, we observed the lattice parameter $a_m/\sqrt{2}$ to be notably larger than that of $c_m$; whereas for x=0.24, we found $a_m/\sqrt{2} < c_m$, where on cooling the value of $a_m/\sqrt{2}$ approached that of $c_m$. Diffraction results then confirmed that these two crystals had identical domain configurations in their low temperature phases, belonging to the $Cm$ space group. Using the values of $a_m/\sqrt{2}c_m$ and $\beta$, we conclude that both PMN-0.15PT and PMN-0.24PT have monoclinic $M_A$ structure (please see Appendix for details).

For PMN-0.27PT and PMN-0.28PT, two phase transitions were observed on electric field cooling with the sequence C→T→$M_A$. In addition, we observed for both crystals that the lattice parameter $a_m/\sqrt{2}$ increased with decreasing temperature, approaching that of $c_m$. The principle difference between the results with increasing PT content in this range was that the field required to stabilize the T phase on cooling was reduced with increasing x: E=0.50kV/cm for x=0.27, but E=0.25kV/cm for x=0.28. For x≥0.3, $M_C$ is the dominant monoclinic phase, as the composition enters the region of the MPB: the transformational sequence for x=0.30 is C→T→$M_C$→$M_A$[15], whereas that for x=0.32 and 0.35 is C→T→$M_C$[16,22]. For x≥0.38, the T phase was stable down to 243K, with a transformational sequence of simply C→T.

Next, we measured changes in mesh scans with increasing x in the [001] FC condition. Figure 4 shows a partial summary of the many measurements made at lower temperatures for various PMN-xPT compositions across the phase diagram. The mesh scans of the low temperature phase with 0.15≤x≤0.30 were consistent with the known signatures for the $M_A$ phase[10,15]; whereas, those with 0.32≤x≤0.35[8,11,15] for the $M_C$ phase. Mesh scans at various temperatures confirmed signatures of the phase transformational sequences noted above for the various compositions.

Our findings for [001] PMN-xPT are summarized in the phase diagram given in Fig. 2(a), above. The [001] FC transformational sequence was C→$M_A$ for x<0.25, C→T→$M_A$ for 0.25≤$x$≤0.3; C→T→$M_C$ for 0.3≤$x$≤0.35; and C→T for $x$>0.35. Based of lattice parameter studies, we found that the intermediate T phase extends to $x$=0.25 in the FC condition, rather than $x$=0.3 as for the ZFC condition. Previous reports of dielectric relaxation[23] in the temperature range of this extended T phase indicate that microdomains of tetragonal symmetry may be stabilized by application of E//[001] over a narrow phase region, on cooling from the cubic to $M_A$ phases. This possibility is further substantiated by the observation that the C→T boundary as determined by structural data ($c>a$) was found to shift to higher temperatures with increasing field for E<3kV/cm, whereas the C'→T boundary determined by the dielectric maximum was independent of E: these results will be discussed in more detail in Section IV.

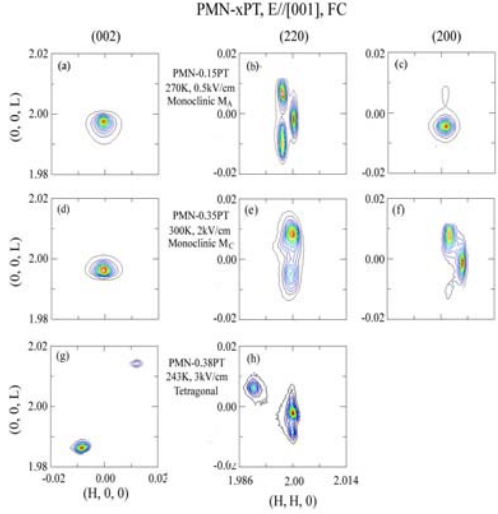

Figure 4. Mesh scans taken about the pseudocubic (002), (220) and (200) reflections for [001] field cooled PMN-$x$PT crystals: (a) PMN-0.15PT, E=0.5kV/cm at 270K; (b) PMN-0.35PT, E=2kV/cm at 300K; and (c) PMN-0.38PT, E=3kV/cm at 243K.

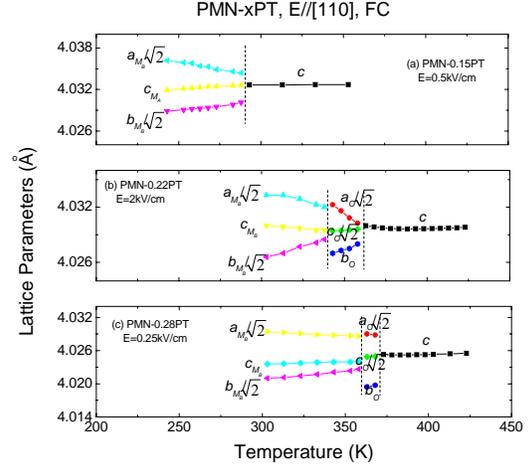

Figure 5. Lattice parameters as a function of temperature for [110] field cooled PMN-$x$PT crystals: (a) PMN-0.15PT, E=0.5kv/cm; (b) PMN-0.22PT, E=2kv/cm; and (c) PMN-0.28PT, E=0.25kv/cm.

### (110) electric field cooled PMN-$x$PT

Figure 5 shows the temperature evolution of the lattice parameter for [110] FC PMN-$x$PT with $x$=0.15, 0.22, and 0.28. Results for PMN-0.30PT[17] and PMN-0.35PT[22] have recently been reported, and can be found in the corresponding references.

For 0.15≤$x$≤0.20, a single phase transition was observed on cooling under E=0.5kV/cm. Structural analysis confirmed that the low temperature phase had a $M_B$ lattice symmetry, with $a_m/\sqrt{2} > c_m$; and thus, the transformational sequence on cooling is C→$M_B$. Temperature dependent lattice parameter studies revealed that the difference between $a_m/\sqrt{2}$ and $c_m$ gradually increased with decreasing temperature in the $M_B$ phase field, contrary to the observed increase in the $M_A$ phase field from the [001] FC measurements. For 0.22<$x$<0.28, two phase transitions were observed on field cooling. Structural analysis revealed the transformational sequence to be C→O→$M_B$: no intermediate T phase was found to extend over to lower PT contents under E, rather a single domain O phase was found in its place. For $x$=0.22, coexistence of O and $M_B$ phases was observed under E≤1.0kV/cm, but with increasing field to E=2kV/cm such coexistence was not found. With increasing x between 0.22 and 0.28, the field required to stabilize the O phase on cooling was found to be decreased. For 0.30≤$x$≤0.35, an intermediate T phase was found, similar to that in the ZFC condition. Over a narrow compositional range near $x$≈0.30, the transformational sequence was found to be C→T→O→$M_B$; whereas for $x$=0.32 and 0.35, we found the sequence to be C→T→O. For $x$>0.32, the T phase was found to become increasingly dominate with increasing $x$; until for $x$=0.38, the sequence was simply C→T.

Next, we also measured changes in mesh scans of PMN-$x$PT ($x$=0.15, 0.35, 0.38) at low temperatures in the [110] FC condition, as shown in Figure 6. Results from one crystal are shown here, which reveals the signatures of both the $M_B$ and O phases. We illustrate the signatures of the $M_B$ phase using results for $x$=0.15 as shown in Figure 6(a), (b), and (c). At 253K, the (220) reflection reveals a single peak, indicating that E//[110] fixes the [110] crystallographic direction of the crystal: two peaks about the (002) reflection were found, which were split along the transverse direction L; and the ($2\bar{2}0$) reflection and (200) reflection (not shown) also revealed a single peak. The combination of these mesh scans provides the signatures of $M_B$-two polarization vectors constrained to the (110) plane that lie close to the [110]. A similar $M_B$ lattice structure was found for $x$=0.22, 0.28, and 0.30[17] at low temperatures. The signature of the O phase are illustrated in Figure 6 (d)-(f) for $x$=0.35, taken at 300K in the FC condition as E=2kV/cm. All mesh scans taken about the (220), (220), (002), and (200) reflections revealed a single peak-demonstrating that a single domain O phase has been induced. For $x$=0.38, the domain configurations as shown in Figure 6 (g) and (h), show that the T phase remains stable on cooling to 243K as

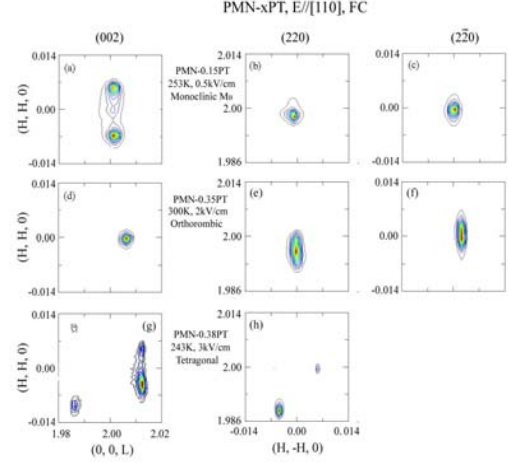

Figure 6. Mesh scans taken about the pseudocubic (002), (220) and (200) reflections for [110] field cooled PMN-$x$PT crystals: (a) PMN-0.15PT, E=0.5kV/cm at 253K; (b) PMN-0.35PT, E=2kV/cm at 300K; and (c) PMN-0.38PT, E=3kV/cm at 243K.

E=3kV/cm.

Our findings for [110] PMN-xPT are summarized in the phase diagram given in Fig. 2(b), above. The [110] FC transformational sequence was C→$M_B$ for $x$≤0.20, C→O→$M_B$ for 0.22≤$x$<0.3; C→T→O→$M_B$ for $x$≈0.30; C→T→O for 0.31<$x$≤0.36; and C→T for $x$>0.37. Based on lattice parameter studies, we found that the intermediate T phase extends to $x$=0.25 in the FC condition, rather than $x$=0.3 as for the ZFC condition. Compared to the [001] FC phase diagram, in the [110] phase diagram, an extended O phase replaces the T and $M_C$ phases, and a $M_B$ phase replaces the $M_A$ phase.

## IV. Discussion

Comparison of the [001] and [110] FC phase diagrams of PMN-$x$PT in Fig. 2 reveals several interesting findings, including: (i) that the R phase of the ZFC state is replaced by $M_A$ in the [001] FC diagram, but with $M_B$ in the [110] FC; (ii) a region (C') of abnormal thermal expansion ($c≠a$) above the dielectric maximum, whose stability range extended to higher temperatures by application of E; (iii) that the $M_C$ phase in the [001] FC diagram is replaced by the O phase in the [110] FC; and (iv) that the stability of the T phase is extended to $x$=0.25 in the [001] FC diagram, whereas this extended T phase region is entirely replaced by the O phase in the [110] FC.

It is important to note that differences between the [001] and [110] phase diagrams was caused by moderate electric fields of 0.25kV/cm≤E≤0.5kV/cm – clearly demonstrating that the phase stability of PMN-$x$PT crystals is quite fragile, or simply put many phases are apparently very close to being energetically degenerate. This phase fragility brings into question conventional wisdoms concerning the thermodynamics of classical phase diagrams – indicating an important role of an underlying structural heterogeneity.

### The $M_A$ and $M_B$ phases

The $M_A$ and $M_B$ phases belong to the same group $Cm$; accordingly, their mesh scans exhibit identical contour features. However, we can distinguish $M_A$ from $M_B$ using the $\beta$ and the ratio of $a_m/\sqrt{2}c_m$ (see appendix for details). Table I shows calculated values of $a_m/\sqrt{2}c_m$ and $\sqrt{2}\cos(\alpha*/2)$ for [001] and [110] FC PMN-$x$PT. Using this table, we can identify the polarization rotation pathway. For [001] FC PMN-15%PT, it was found that $a_m/\sqrt{2} > c_m$ over the entire phase field of the low temperature phase. Thus, we can conclude that the polarization is constrained to the R→T path of the $M_A$ phase, and it is not possible that it would follow the R→O one of $M_B$. However, for [110] FC PMN-$x$PT with 0.15<$x$<0.3, it was found that $a_m/\sqrt{2}c_m > \sqrt{2}\cos(\alpha*/2)$ over the entire phase field of the low temperature phase, demonstrating that polarization rotation occurs towards [110] away from [111], following the path R→O of the $M_B$ phase.

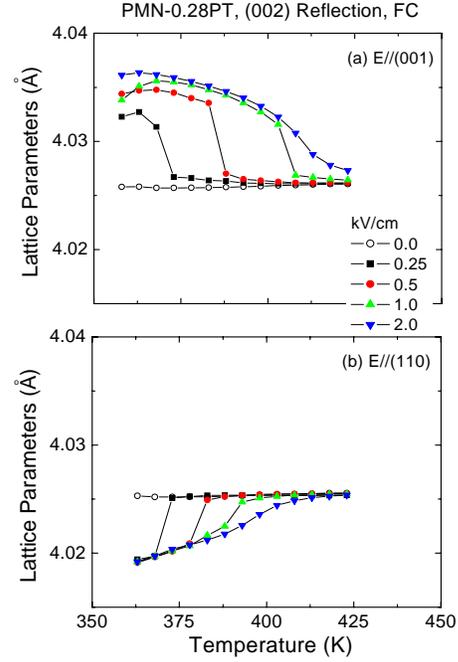

Figure 7. Lattice parameter as a function of temperature derived from the (002)-reflection of PMN-0.28PT under different electric field levels applied along: (a) E//[001], and (b) E//[110].

We note several other important points. For $x$=0.15, the difference between $a_m/\sqrt{2}c_m$ and $\sqrt{2}\cos(\alpha*/2)$ is very small, demonstrating that the polarization vector lies quite close to [111]. Second for [001] FC PMN-$x$PT with 0.24≤$x$≤0.27, we observed that the value of $a_m/\sqrt{2}$ approaches that of $c_m$ with decreasing temperature. This demonstrates that the polarization of the $M_A$ phase gradually rotates back on cooling towards [111] away from the [001].

### The C' phase

Both the [001] and [110] phase diagrams of PMN-$x$PT exhibited a region of abnormal thermal expansion, designated as C'. In this region the lattice parameter *c* derived from the (002) reflection was found not to be equal to that of *a* derived from the (200) reflection. In addition, it was observed that the stability of the C' phase was extended to higher temperatures following application of E. As an example, Figure 7 shows the temperature dependence of the lattice parameters derived from the (002) reflection for PMN-0.28PT in the FC state at various fields for (a) E//[001], and (b) E//[110]. Both figures show that the region of $c/a>1$ extends to higher temperatures with increasing E. Please note for E=2kV/cm that the lattice parameters changed continuously with decreasing temperature, rather than exhibiting an abrupt anomaly as for E=0.25kV/cm.

We are careful to distinguish the boundary C' from that of the true cubic C: C was determined from the Curie temperature ($T_C$) of the dielectric maximum, whereas C' from the temperature at which $c/a$ first deviated from one. Thus, in the phase field C', we have tetragonal splitting that occurs above the temperature of the dielectric maximum. This is unconventional with respect to normal phase transitions: where symmetry breaking occurs at or below $T_C$, but never above. Rather, it has some similarities to relaxor ferroelectric behavior in PMN, where local polarizations are known to exist above the dielectric maximum. Accordingly, we attribute this anomalous phase field C' to polar nano-regions (PNRs). However, there are important differences between the phase field C' and a relaxor state. Relaxors are well-known to have a pseudo-cubic structure, averaging over PNRs that are randomly distributed amongst all possible domain variants on a nanometer length scale yields a structure that appears cubic on average. However, in the phase field C', under E//[001] or E//[110], the PNRs will be partially aligned under the direction of the applied E. Thus, the local structural asymetric distortions of the PNRs will not be averaged to cubic, rather an ensemble of PNRs will exhibit a net distortion. Accordingly, a symmetry breaking above $T_C$ may be observed, which gradually changes with E and temperature as the topological arrangement of PNRs is altered.

Table I Calculated value of $a_m/\sqrt{2}$ and $\sqrt{2}\cos(\alpha^*/2)$ for low temperature phases in [001] and [110] electric field cooled PMN-*x*PT crystals. $\beta$, $a_m/\sqrt{2}$ and $c_m$ were directly derived from the experiments and the imaginary rhombohedral angle $\alpha^*$ was determined according to $\cos(\beta)=(1-2\sin^2(\alpha^*/2))/\cos(\alpha^*/2)$.

| x | T(K) | E(kV/cm) | $a_m/\sqrt{2}$ (Å) | $c_m$(Å) | $\beta$(°) | $\alpha^*$(°) | $a_m/(\sqrt{2}c_m)$ | $\sqrt{2}\cos(\alpha^*/2)$ | $a_m/2c_m\cos(\alpha^*/2)$ | S |
|---|---|---|---|---|---|---|---|---|---|---|
| | | | | | E//[001] | | | | | |
| 0.30 | 300 | 1.0 | 4.0245 | 4.0242 | 89.950 | 89.965 | 1.0001 | 1.0003 | 0.9998 | $M_A$ |
| 0.28 | 353 | 0.25 | 4.0258 | 4.0314 | 89.895 | 89.965 | 0.9986 | 1.0007 | 0.9979 | $M_A$ |
| 0.27 | 300 | 0.35 | 4.0299 | 4.0280 | 89.850 | 89.894 | 1.0005 | 1.0010 | 0.9995 | $M_A$ |
| 0.24 | 300 | 0.5 | 4.0295 | 4.0302 | 89.870 | 89.908 | 0.9998 | 1.0008 | 0.9990 | $M_A$ |
| 0.15 | 270 | 0.5 | 4.0348 | 4.0319 | 89.885 | 89.917 | 1.0007 | 1.0008 | 0.9999 | $M_A$ |
| | | | | | E//[110] | | | | | |
| 0.30 | 300 | 1.0 | 4.0280 | 4.0200 | 89.850 | 89.894 | 1.0020 | 1.0010 | 1.0009 | $M_B$ |
| 0.28 | 353 | 0.25 | 4.0287 | 4.0240 | 89.900 | 89.929 | 1.0012 | 1.0006 | 1.0006 | $M_B$ |
| 0.22 | 333 | 2.0 | 4.0322 | 4.0295 | 89.920 | 89.945 | 1.0007 | 1.0005 | 1.0002 | $M_B$ |
| 0.15 | 243 | 0.5 | 4.0368 | 4.0319 | 89.853 | 89.897 | 1.0012 | 1.0009 | 1.0003 | $M_B$ |

Here, we consider the structural evolution from the cubic to ferroelectric state to occur in three steps. First, near $T_B$ (Burns temperature), clusters of short-range polar order (i.e. PNR's) develop and gradually increase in number on cooling. Second, with decreasing temperature, the ensemble of PNR's becomes percolating, resulting in the development of microdomains. However, since these micro-domains are on a scale much smaller than the x-ray coherence length, only a slight broadening of the Brags reflections is observed. This evolution is believed to be continuous and the dielectric behavior correspondingly varies smoothly on cooling. Third, below a temperature that we designate as $T_M$, a thermally activated formation of critical ferroelectric nuclei occurs from the microdomain state – below which point, a long-range ordered ferroelectric phase is stable. Application of field assists in aligning the PNR along the direction that E is applied, resulting in PNR growth and an increase in its numbers, giving rise to the abnormal thermal expansion above $T_C$ that defines the C' phase field. Higher fields favor PNR growth at higher temperature; and correspondingly, the C'→C phase boundary is shifted to higher temperature with increasing E.

It is worth noting that the stability range of the C' phase field was extremely narrow when the stable low temperature phase was tetragonal in both the [001] and [110] phase diagrams. The C' phase field widens significantly with increasing E only in the region where the transformational sequence was altered by changing the direction along which E was applied. The most pronounced changes were found for $x \approx 0.30$ where the sequence was C→C'→T→$M_C$→$M_A$ for E//[001], but C→C'→T→O→$M_B$ for E//[110]. These observations suggest that the fragileness of the phase stability might be related to the structural inhomogeneity originating from the PNR state. Recent experiments on PZN-xPT relaxor systems suggest that an external electric field along [111] direction actually enhances PNR with polarizations perpendicular to, instead of along the external field in the rhombohedral phase. Although the implications of these results on our [001] and [110] electric field measurements are not yet clear, they do indicate that the field can greatly affect the PNR configuration, and therefore the structural inhomogeneity [24,25].

**Extension of T and/or O phase fields to lower *x* by E.**

In the ZFC phase diagram of PMN-*x*PT, the T phase extends only to $x=0.30$; and for $x<0.30$, a C→R sequence is found on cooling. For the [001] FC crystals, the T phase was extended to $x=0.25$, and could be induced by fields as small as E=0.25kV/cm for $x=0.28$. However, for the [110] FC crystals, the region where the T phase was extended was entirely replaced by the O phase.

Following the phase diagrams in Figure 2, it can be noted that the $M_C$/O phases are closely related to the presence of a stable T phase, whereas $M_A$/$M_B$ are always connected to R. Thus, we can conclude that for compositions on the left side of the MPB ($x<0.30$), the polarization rotation pathway is R→$M_A$→T for E//[001] and R→$M_B$→O for E//[110] with increasing E beginning from the ZFC condition; and that for compositions inside of the MPB ($0.30<x\leq0.35$), the path is $M_C$→T for E//[001], but $M_C$→O for E//[110]. A special case occurs for PMN-0.30PT: the rotational pathway is R→M$_A$→M$_C$→T with increasing E from the ZFC, due to a monoclinic M$_A$ to M$_C$ transition.

In the PMN-*x*PT crystalline solution, the substitution of the octahedron [TiO$_4$]$^{4-}$ for the more complex [Mg$_{1/2}$Nb$_{2/3}$O$_4$]$^{4-}$ one, between relaxor PMN and ferroelectric PT, results in a MPB separating ferroelectric R and T phase. In the T phase field, the tetragonal splitting is known to be weakened with decreasing $x^{14,26}$. For *x*=0.30 in the ZFC condition, the T phase is only observed over a narrow range of temperatures; furthermore, relaxor ferroelectric behavior has been reported in ZFC crystals for *x*≤0.30. Thus, PMN-0.30PT can be considered as a special composition in the phase diagram, where a gradual transition between microdomains of the T phase and a macroscopic T phase begins to occur. We note that microdomains have been observed by electron microscopy for 0<*x*<0.30$^{27}$. Since these microdomains are much smaller than the coherence length of x-rays, the structure appears cubic below $T_M$ in the ZFC state. However, in the FC state, the microdomains align along the direction that E is applied. Thus, an extended macroscopic T phase is observed to be sandwiched between the C and M$_A$ ones on cooling for 0.25≤*x*<0.30, whose *c/a* ratio can be significantly altered by E. However, more thought is required to understand the presence of microdomains and intermediate orthorhombic and monoclinic phases.

**Fragile phase stability of PMN-*x*PT**

Consider for example PMN-0.28PT, T and O phases were found in the same temperature range for E//[001] and E//[110] respectively, under fields as small as *E=0.25kV/cm*. However, comparison of the lattice parameters of the T and O phases (at the same temperature) reveals an interesting feature: $a_T \approx b_O \approx 4.020$Å and $a_o/\sqrt{2} + c_o/\sqrt{2} = a_T + c_T$ ≈8.049Å (365K). Similar observations were made at other temperatures and for *x*=0.30$^{17}$, 0.35$^{22}$. Since the O phase is the limiting case of the M$_C$ one, we observed an equally important relationship between the lattice parameters of the T and M$_C$ phases for x=0.30$^{17}$, 0.32$^{16}$, 0.35$^{22}$: $b_M = a_T$, and $a_M + c_M = a_T + c_T$ where $a_M$, $b_M$, and $c_M$ are the lattice parameters of the M$_C$ unit cell. These observations demonstrate the existence of an important crystallographic relationship/transformation between the T, O, and M$_C$ phases – they are not independent of each other.

Recently, a theory of an adaptive ferroelectric phase$^{28-30}$ has been developed to predict the microdomain-averaged crystal lattice parameters of a structurally inhomogeneous state, consisting of tetragonal microdomains. This theory predicts crystallographic relationships between T and M$_C$ phases of

$$a_M + c_M = a_T + c_T \quad (1a)$$

$$b_M = a_T; \quad (1b)$$

which have been experimental verified for PMN-*x*PT and PZN-*x*PT. Recently, Wang *et al.*$^{31}$ have extended this analysis, obtaining a relationship between the monoclinic angle ($\beta$) and the tetragonal/monoclinic lattice parameters, given as

$$\beta = 90° + 2A\omega(1-\omega)\left(\tan^{-1}\frac{c_T}{a_T} - 45°\right) \quad (1c)$$

where $\omega = \dfrac{c_M - b_M}{a_M + c_M - 2b_M}$, $a_T = b_M$, $c_T = a_M + c_M - b_M$, and the constant $A \approx 1$. Furthermore, we note that the O phase is a limiting case of M$_C$, where $\omega = ½$.

The predictions of (1) are identical to the experimentally observed relationship between the T, O, and $M_C$ lattice parameters that we noted above. These observations provide quantitative evidence that the $M_C$ and O phases are adaptive phases consisting of tetragonal microdomains. Application of E//[110] fixes the [110] orientation, where the [100] and [001] variants of tetragonal microdomains are of equal volume fraction: thus, the stable phase appears to be a single domain orthorhombic. Whereas for E//[001], the volume fraction of the tetragonal microdomains variants are variable and not equivalent: thus, the stable phase appears to be polydomain $M_C$.

In summary, the results of this investigation demonstrate that the phase stability is "fragile": the phase diagrams can be altered by application of modest electric fields along different crystallographic axis. However, comparisons of the lattice parameters of the different phase fields shows that the "fragility" may only be a perception – the lattice parameters of the O, $M_C$, and T phase are inter related. Analysis of the lattice parameters suggests that in fact the O and $M_C$ phases consist of tetragonal microdomains that are geometrically aligned with respect to each other, in manner to achieve stress accommodation. In this case, the O and $M_C$ phases may appear to be uniform on a length equivalent or larger than that of the coherence length of x-rays, but in fact consists of a structurally inhomogeneous phases of tetragonal microdomains on a local scale.

## Acknowledgements

We would like to gratefully acknowledge financial support from the Office of Naval Research under grants N000140210340, N000140210126 and MURI N0000140110761; U.S. Department of Energy under Contract No. DE-AC02-98CH10886. We would like to thank HC Materias for providing the single crystals used in this study.

## VI. Appendix

Although the existence of $M_A$ and $M_B$ phases in PMN-$x$PT has long been predicted by Vanderbilt and Cohen using a thermodynamic theory[19], it is not easy to distinguish these two phases in experiments. For $M_A$ phase, $Px=Py<Pz$; where as for $M_B$ phase, $Px=Py>Pz$. However, in structural measurements, comparing $Px$, $Py$, and $Pz$ is not always straight-forward. Here we discuss a robust but easy-to-use criterion that can be applied to structural measurements.

A schematic for the $M_A/M_B$ phases is shown in Fig. 8. Here we show the plane defined by the [110] and [001] vectors. In both monoclinic phases, [110] ($a_m$) is tilted up toward [001] ($c_m$), and the length of $a_m$ and $c_m$ also deviate from those in the cubic phase (in the cubic phase, $a_m = \sqrt{2}a = \sqrt{2}c$). Therefore, the ratio $a_m/\sqrt{2}c_m$ and the monoclinic angle $β$ between $a_m$ and $c_m$ are the only two parameters necessary to define a $M_A/M_B$ phase.

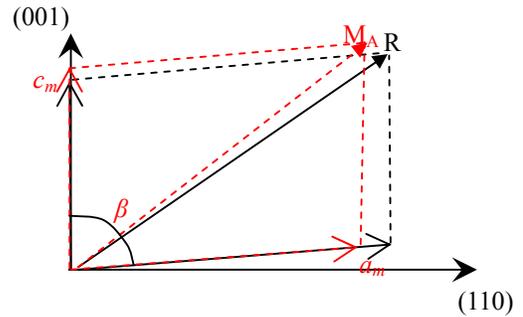

Figure 8. Illustration of polarization vectors of the rhombohedral (R) and monoclinc $M_A$ ($Cm$) phases, represented with same monoclinic angle $β$ in the HHL zone of reciprocal space. The R unit cell can be expressed in terms of a monoclinic one by: $a_m=2a_r\cos(α/2)$, $b_m=2a_r\sin(α/2)$, $a_m=a_r$, $\cos(β)=(1-2\sin^2(α/2))/\con(α/2)$, where $a_r$ and $α$ are the $R3m$ cell parameters.

For a monoclinic phase with a fixed monoclinic angle $\beta$, the polarization components ($P_x, P_y, P_z$) are exclusively determined by the ratio $a_m/\sqrt{2}c_m$. The phase is $M_A$ if the polarization falls closer to the [001] direction (T), $M_B$ if the polarization falls closer to the [110] direction (O). The severing point between $M_A$ and $M_B$, is therefore the phase where $P_x=P_y=P_z$, which is a rhombohedral phase (R).

This provides a definitive way to distinguish $M_A$ and $M_B$ based on $\beta$ has to satisfy $\cos(\beta)=(1-2\sin^2(\alpha*/2))/\cos(\alpha*/2)$ for the angle between [001] and [110] to be $\beta$). If $a_m/\sqrt{2}c_m$ from our measurements is greater than $\sqrt{2}\cos(\alpha*/2)$, we have a $M_B$ phase, otherwise we have a $M_A$ phase. Since for most cases in lead perovskite relaxors, $\alpha$ and $\beta$ are very close to 90° degrees, $\sqrt{2}\cos(\alpha*/2)$ is and $a_m/\sqrt{2}c_m$. When we have a $M_A/M_B$ type structure, we should compare its $a_m/\sqrt{2}c_m$ ratio to that of the rhombohedral phase which has the same angle $\beta$ between its [001] and [110] vectors. For example, for a rhombohedral phase with rhombohedral angle $\alpha*$ (angle between [001] and [100]), the ratio $a_m^*/\sqrt{2}c_m^*$ (here $a_m^*$ denotes the length along [110] direction) is $\sqrt{2}\cos(\alpha*/2)$, while $\alpha*$ also close to 1. In practice, people usually can compare $a_m/\sqrt{2}c_m$ to 1 to determine whether the structure is a $M_A$ or $M_B$ phase[15,17,26]. However, when $a_m/\sqrt{2}c_m$ is close to 1, the method described here should be used.

_______________________________________________